\documentclass[12pt]{article}
\usepackage[margin=25mm]{geometry}
\usepackage{amsmath}
\usepackage{graphicx,psfrag,epsf}
\usepackage{enumerate}
\usepackage{natbib}
\usepackage{amssymb}
\usepackage{epstopdf}
\usepackage{enumerate}
\usepackage{placeins}
\usepackage{setspace}
\usepackage{float}
\usepackage{soul,color}
\usepackage{placeins}
\usepackage{upgreek, mathtools,bm}
\usepackage{algorithm,algpseudocode,tabularx}
\usepackage{caption}
\usepackage{subcaption}
\usepackage{enumitem}
\usepackage{multirow}
\usepackage{amsfonts,amssymb,graphics,amsthm, comment}
\usepackage{url}
\RequirePackage[colorlinks,breaklinks,linkcolor=blue,citecolor=blue,urlcolor=blue]{hyperref}
\usepackage{multicol}
\usepackage{listings} 
\usepackage{xcolor} 
\usepackage{siunitx}
\immediate\write18{texcount -tex -sum  \jobname.tex > \jobname.wordcount.tex}

\newcommand{\N}{\mathbb{N}}
\newcommand{\eps}{\varepsilon}
\newcommand{\E}{\mathbb{E}}
\newcommand{\T}{\mathcal{T}}
\newcommand{\R}{\mathbb{R}}
\newcommand{\B}{\mathcal{B}}
\renewcommand{\P}{\mathcal{P}}

\def\bA{\boldsymbol{A}}
\def\bB{\boldsymbol{B}}

\def\bG{\boldsymbol{G}}
\def\bP{\boldsymbol{P}}

\def\bT{\boldsymbol{T}}

\newcommand{\Lp}{\mathcal{L}}
\newcommand{\hs}{\mathcal{H}}
\newcommand{\K}{\mathcal{K}}
\newcommand{\I}{\mathcal{I}}

\newcommand{\amax}{\arg\max}
\newcommand{\amin}{\arg\min}
\newcommand{\what}[1]{\widehat{#1}}
\newcommand{\wtilde}[1]{\widetilde{#1}}
\renewcommand\d[1]{\:\textrm{d}#1}
\newcommand{\idfun}{\mathrm{id}}

\DeclarePairedDelimiterX{\norm}[1]{\lVert}{\rVert}{#1}
\DeclarePairedDelimiterX{\normHS}[1]{\lVert}{\rVert_{\text{HS}}}{#1}
\DeclarePairedDelimiterX{\normOP}[1]{\lVert}{\rVert_{\text{op}}}{#1}
\DeclarePairedDelimiterX{\abs}[1]{\lvert}{\rvert}{#1}
\newcommand{\innerproduct}[2]{\left\langle #1, #2 \right\rangle}
\newcommand{\innerproductHS}[2]{\left\langle #1, #2 \right\rangle_{\text{HS}}}

\DeclareMathOperator{\rank}{rank}

\DeclarePairedDelimiterX{\sbrac}[1]{[}{]}{#1}
\DeclarePairedDelimiterX{\pbrac}[1]{(}{)}{#1}

\def\hpsi{\hat{\psi}}

\def\tphi{\tilde{\phi}}

\newtheorem{thm}{Theorem}[section]

\theoremstyle{remark}

\newtheorem*{example}{Example}

\theoremstyle{plain}
\theoremstyle{definition}
\newtheorem{remark}{Remark}
\newtheorem{assp}{Assumption}
\providecommand{\keywords}[1]
{
  \small	
  \textbf{\textit{Keywords---}} #1
}
\title{Learning Shared and Source-specific Subspaces\\
across Multiple Data Sources for Functional Data}
\author{Chi Zhang, Peijun Sang, Yingli Qin\\
        \small Department of Statistics and Actuarial Science, University of Waterloo \\
}
\date{}

\begin{document}
\maketitle

\begin{abstract}
In the era of big data, integrating multi-source functional data to extract a subspace that captures the shared subspace across sources has attracted considerable attention. In practice, data collection procedures often follow source-specific protocols. Directly averaging sample covariance operators across sources implicitly assumes homogeneity, which may bias the recovery of both shared and source-specific variation patterns. To address this issue, we propose a projection-based data integration method that explicitly separates the shared and source-specific subspaces. The method first estimates source-specific projection operators via smoothing to accommodate the nonparametric nature of functional data. The shared subspace is then isolated by examining the eigenvalues of the averaged projection operator across all sources.  If a source-specific subspace is of interest, we re-project the associated source-specific covariance estimator onto the subspace orthogonal to the estimated shared subspace, and estimate the source-specific subspace from the resulting projection. We further establish the asymptotic properties of both the shared and source-specific subspace estimators. Extensive simulation studies demonstrate the effectiveness of the proposed method across a wide range of settings. Finally, we illustrate its practical utility with an example of air pollutant data.
\end{abstract} \hspace{10pt}

\section{Introduction}
Functional data analysis (FDA) has gained increasing prominence in modern statistics, driven by advances in data collection technologies. It provides a nonparametric framework for analyzing discrete observations taken from realizations of a continuous random function, often defined over time or space. The inherently infinite-dimensional nature of functional data necessitates the use of dimension reduction techniques, which transform infinite-dimensional random functions into finite-dimensional random vectors and thereby facilitate subsequent analysis. Among these techniques, functional principal component analysis (FPCA) is one of the most widely used; see, for example, Chapter 8 of \citet{ramsayFDA2005} or Chapter 11 of \citet{kokoszka_introduction_2017}. Analogous to principal component analysis, FPCA aims to extract a parsimonious subspace that explains the most relevant variation around a mean function in a data-adaptive manner.

The theoretical foundations of FPCA have been extensively developed over the past two decades \citep{yao_functional_2005, hall_properties_2006, zhou_theory_2024}. FPCA is also commonly adopted as a regularization tool when inverting the covariance operator is required. For example, the inverse of the covariance operator is unbounded without regularization in functional linear regression \citep{hall_methodology_2007, zhou_functional_2023} and functional generalized linear models \citep{dou_estimation_2012}. While FPCA methods are well established both in theory and applications, comparatively less attention has been devoted to settings involving multiple functional data sources, where one seeks to aggregate information and uncover shared structure. Yet, such scenarios are increasingly common in modern applications, where data collected at different sites often present heterogeneity. For instance, in the air pollutants study discussed in Section \ref{sec:real_data}, measurements are obtained from five cities with distinct geographical locations and climate conditions.

\begin{figure}[!ht]
    \centering
    \includegraphics[width=\textwidth, height = 20em]{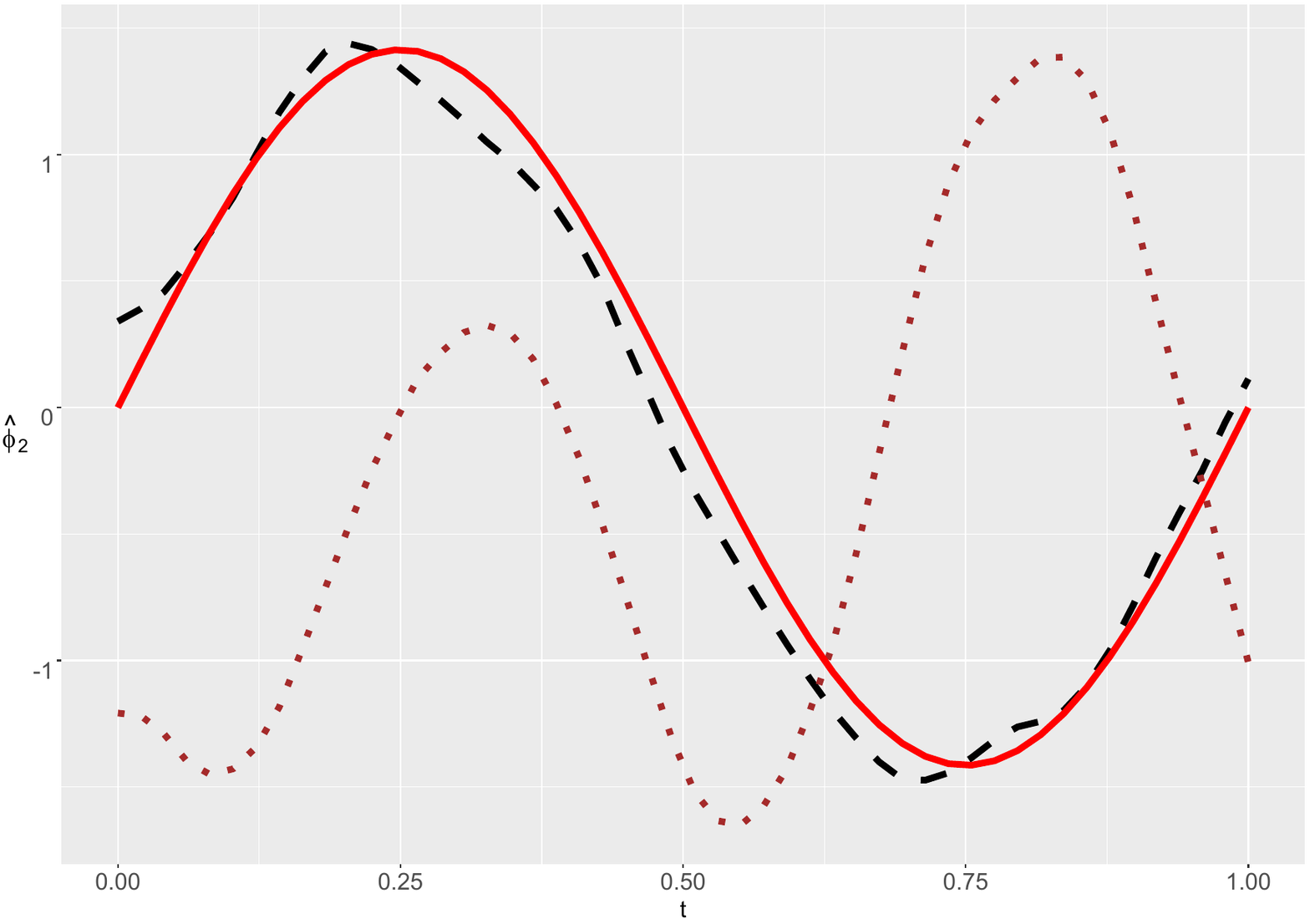}
    \caption{Data are generated from model \eqref{eq:model} and Example \ref{example:Fourier_rotation}. The solid line represents the true shared eigenfunction across two sites. The dashed line shows the estimated shared eigenfunction obtained using the proposed method in Section \ref{sec:multiFPCA}. The dotted line corresponds to the eigenfunction estimated from the averaged covariance operator.}
    \label{fig:example1_phi2_comparison}
\end{figure}

When functional data are collected from multiple sources, heterogeneity across sources poses additional challenges for analysis. A naive extension of single-source FPCA is to compute source-specific sample covariance operators and then perform FPCA on their average. However, this averaging procedure implicitly assumes homogeneity across sources, and dominant patterns present only in particular sources may introduce severe bias in identifying the true shared structure. To illustrate, we construct a toy example in Example \ref{example:Fourier_rotation} with two sources. The covariance operator of the second source is generated by applying a simple rotation to the basis functions of the first source, while retaining the same eigenvalues. Even though the two sources have identical sample sizes and eigenvalues, the eigenfunction estimated from the averaged covariance operator deviates substantially from the truth, as shown in Figure \ref{fig:example1_phi2_comparison}.

Beyond naive averaging, much of the existing literature on multi-source functional data has focused on mean estimation under the framework of transfer learning. For example, \citet{cai_transfer_2024} studied phase transition phenomena in mean function estimation based on discretely sampled functional data and established the corresponding minimax rates. Another line of work concerns common FPCA (CFPCA) and partially common FPCA (PCFPCA). CFPCA assumes that the leading eigenfunctions for all sources are the same, with differences arising only in the eigenvalues \citep{BenkoCFPC2007AOS, BOENTE2010464}, whereas PCFPCA allows some eigenfunctions to be shared and others to remain source-specific \citep{jiao_filtrated_2024}. For instance, \citet{jiao_filtrated_2024} considered a forest-structured hierarchy of communities, where sources in each layer were partitioned into communities such that the data sources within the same community partially shared a subset of basis functions.

However, existing CFPCA and PCFPCA methods are designed for fully or densely observed functional data, which may be too restrictive in many real-world applications. Moreover, although the parameter of interest is often a low-dimensional shared subspace, most existing approaches rely on linear combinations of covariance functions estimated from each source. Such strategies remain prone to bias when the eigenvalues associated with source-specific directions dominate those of the shared subspace across sources, regardless of how the averaging weights are chosen. A more natural perspective is to focus directly on the subspace itself: if the shared subspace is the primary object of interest, one can safely discard the eigenvalues during the data integration process and retain only the eigenfunctions. This perspective motivates the projection-based approach we develop in this paper.

To our knowledge, no existing methods directly pursue this subspace-focused strategy in the functional data literature to extract the shared subspace. The idea is inspired by \cite{li_knowledge_2024}, which aims to enhance the estimation efficiency of the shared structure under the framework of transfer learning for multivariate data. They identified the shared space by solving an optimization problem on the Grassmann manifold, where the manifold is often defined for finite-dimensional Euclidean space.

We generalize the method to functional data, where the underlying space is infinite-dimensional. Specifically, we first estimate the sample covariance function for each source using a local linear smoother, thereby accommodating the nonparametric nature of functional data. Aggregating observations within each source allows us to borrow strength across subjects, enabling estimation even for sparse functional data. Discretizing the smoothed covariance estimates yields estimated eigenpairs and corresponding sample projection operators for each source. We then construct a weighted average projection estimator, $\what{\bP}_w$, from the source-wise sample projection operators. By utilizing the spectral properties of $\what{\bP}_w$, we identify the shared subspace, which is then used to estimate the source-specific subspace by re-projecting the estimated covariance operator of the specific source onto the orthogonal complement of the shared subspace. 

Furthermore, we establish theoretical convergence guarantees for both the shared and source-specific estimators. The convergence rate of the estimator of the shared projection operator highlights the gain in estimation efficiency achieved by effectively leveraging information from multiple data sources. These theoretical insights align with the numerical results from simulation studies under various settings, where the proposed method consistently outperforms single-source estimators.

The remainder of this article is organized as follows. In Section \ref{sec:standard_FPCA}, we introduce the data generation mechanism and standard FPCA for one source. Then, we rigorously define the shared structure for functional data in Section \ref{sec:shared_structure}. The proposed method is detailed in Section \ref{sec:multiFPCA}. We develop asymptotic results for the proposed estimator in Section \ref{sec:theory}. Detailed proofs are relegated to the online supplement. We showcase the finite-sample performance of the proposed method under various simulation settings in Section \ref{sec:simulation}. In Section \ref{sec:real_data} we apply the proposed method to a real world example. 
Finally, we conclude the paper in Section \ref{sec:conclusions}.

\section{Estimation via Data Integration}\label{sec:est}
\subsection{Standard FPCA}\label{sec:standard_FPCA}
Before presenting the proposed method for multiple data sources, we begin by reviewing the standard FPCA for a single source and introducing the notation. Let $X(t)$ be a continuous and square-integrable stochastic process defined on a compact interval $\T = [0,1]$ with $\E[X(t)] = 0$ for $t \in \T$. Denote $\hs = \Lp^{2}(\T)$, the Hilbert space of square-integrable functions on $\T$, equipped with inner product $\innerproduct{f}{g}_2 = \int_{t \in \T}f(t)g(t)\d t$. Moreover, $\norm{f}_2^2 = \innerproduct{f}{f}_2$. For a fixed $m$, FPCA seeks an $m$-dimensional subspace $\hs_0 \subset \hs$ such that the projection of $X$ onto $\hs_0$ provides the optimal mean-squared approximation to $X$. More precisely, FPCA identifies $m$ orthonormal functions $e_1, e_2, \ldots, e_m \in \hs$ that minimize the expected mean-squared distance between $X$ and its $m$-dimensional projection, which is formalized as follows:
\begin{equation}\label{eq:PCA_projection_expression}
    \wtilde{\bP} = \underset{\bP \in \P_m}{\amin}~\E \norm{X - \bP X}_2^2.
\end{equation}
Here, $\P_m = \{\bP \in \B(\hs) \mid \bP^2 = \bP, \rank(\bP) = m,\bP \text{ is self-adjoint}\}$, and $\B(\hs)$ denotes the set of bounded linear operators on $\hs$. In addition, any $\bP \in \P_m$ can be written as $\bP = \sum_{\nu = 1}^{m}e_\nu \otimes e_{\nu}$, where $f \otimes g(\cdot) = \innerproduct{f}{\cdot}g$ is a rank-one operator on $\hs$ for $f, g \in \hs$.  

For two operators $\bA, \bB \in \B(\hs)$, define $\innerproductHS{\bA}{\bB} = \sum_{\nu}\innerproduct{\bA(e_{\nu})}{\bB(e_{\nu})}_2,$ where $\{e_\nu:\nu =1,2,\ldots\}$ is any orthonormal basis of $\hs$. This definition does not depend on the choice of basis and induces the Hilbert–Schmidt norm $\normHS{\bA}^2 = \innerproductHS{\bA}{\bA}$. With this notation, the optimization problem in \eqref{eq:PCA_projection_expression} is equivalent to
\begin{equation}\label{eq:PCA_projection_expression_alternative}
    \wtilde{\bP} = \underset{\bP \in \P_m}{\amin}~\E \norm{X - \bP X}_2^2 - \E \norm{X}_2^2= \underset{\bP \in \P_m}{\amax} \innerproductHS{\bG}{\bP},
\end{equation}
where $\bG$ is the covariance operator induced by the covariance function of $X$. Specifically, for any $f \in \hs$,
\begin{equation*}
    (\bG f)(t) = \int_{s \in \T}G(s,t)f(s)\d s
\end{equation*}
with $G(s,t) = \E [X(s)X(t)]$ for $s, t \in \T$. The continuity of $X$ together with the definition of $G$ implies that $G$ is continuous, symmetric, and positive semi-definite. By Mercer’s theorem,
\begin{equation}\label{eq:Mercer_thm}
    G(s,t) = \sum_{\nu=1}^{\infty}\lambda_\nu\phi_\nu(s)\phi_\nu(t), \quad t,s \in \T,
\end{equation}
where $(\lambda_\nu, \phi_\nu)$ is the $\nu$th eigenvalue-eigenfunction pair of $\bG$ satisfying $\bG\phi_{\nu} = \lambda_{\nu} \phi_{\nu}$, and $\{\phi_{\nu}\}_{\nu \geq 1}$ forms an orthonormal basis of $\hs$. Combining equations \eqref{eq:PCA_projection_expression_alternative} and \eqref{eq:Mercer_thm}, it follows immediately that $e_i = \phi_i$ for $i = 1, \ldots, m.$ Therefore, it implies that we can use the leading $m$ eigenfunctions of $\bG$ to approximate $X$.

In practice, we can hardly get access to fully observed sample paths without any measurement error for $X$. Suppose that there are $n$ i.i.d. sample paths of $X$, with $N_i$ observations for subject $i$, observed at times $T_{ij}$ with additive measurement error:
\begin{equation*}
    Y_{ij} = X_{i}\left(T_{ij}\right) + \eps_{ij}, \quad i=1, 2, \ldots, n,~j=1, 2, \ldots, N_i.
\end{equation*}
Here $N_i$ are i.i.d.\ copies of a random variable $N$; ${T_{ij}}$ are subject-specific observation times in $\T$; and $\eps_{ij}$ are independent with mean zero and variance $\sigma_\eps^2$. Moreover, $N$, $T$, $\eps$, and $X$ are assumed to be independent. This model is widely adopted in modeling longitudinal observations under the framework of functional data; see \cite{yao_functional_2005}, \cite{li_uniform_2010}, \cite{zhang_sparse_2016}, \cite{zhou_theory_2024} and references therein.

We first estimate the covariance function ${G}(t,s)$ by pooling observations across subjects, following the idea in \citet{yao_functional_2005}, where a local linear smoother is employed. Next, we estimate the eigenvalue–eigenfunction pairs $(\lambda_\nu, \phi_\nu), \nu = 1,2,\ldots,m,$ from the discretization of the smoothed covariance function; see \citet{rice_estimating_1991} and \cite{ramsayFDA2005} for more details. The choice of $m$ can be determined by the fraction of variance explained (FVE), information-criterion-based methods \citep{yao_functional_2005, li_selecting_2013}, or methods integrating information from both eigenvalues and eigenfunctions \citep{zhang2025determineorderfda}.

\subsection{Multi-source Model}\label{sec:shared_structure}
Suppose that there are $K$ sources, and let $\K = \{1, 2, \ldots, K\}$ denote the index set. For any $k \in \K$, we assume that the observations $Y_{ij}^{[k]}$ collected at $T_{ij}^{[k]}$ are generated from the following:
\begin{equation}\label{eq:model}
    Y_{ij}^{[k]} = X_{i}^{[k]}\left(T_{ij}^{[k]}\right) + \eps^{[k]}_{ij}, \quad i=1, 2, \ldots, n_k,~j=1, 2, \ldots, N_i^{[k]}.
\end{equation}
Analogous to the single-source model, we assume $N_i^{[k]}$ are i.i.d. copies of $N^{[k]}$, and $T_{ij}^{[k]}$'s are observation times in $\T$ for source $k$. Measurement errors $\eps_{ij}^{[k]}$ are i.i.d with mean zero and variance $(\sigma_{\eps}^{[k]})^2$. Moreover, $N^{[k]}$, $T^{[k]}$, $\eps^{[k]}$ and $X^{[k]}$ are independent. 

The Karhunen–Lo\`{e}ve expansion of $X^{[k]}(t)$ for $k = 1, \ldots, K,$ is
\begin{equation}\label{eq:KL_expansion}
    X^{[k]}(t) = \sum_{\nu=1}^{\infty}\xi_\nu^{[k]}\phi_{\nu}^{[k]}(t), ~~t \in \T,
\end{equation}
where $\xi_\nu^{[k]} = \int_{\T} X^{[k]}(t)\phi^{[k]}_\nu(t) \d t$ are uncorrelated mean-zero random variables with variance $\lambda_\nu^{[k]}$. It is common to assume that the underlying process $X^{[k]}$ is well-approximated by the leading $m_k$ eigenfunctions \citep{yao_functional_2005}. Moreover, the number of eigenfunctions that can be consistently estimated depends on the sample size and the number of observations per subject \citep{zhou_theory_2024}. Throughout this paper, we focus on the leading $m_k$ functional principal components for source $k$. More precisely, we assume $\rank(\bP_k) = m_k$ for $k \in \K$. This assumption is standard in the CFPCA/PCFPCA literature, where $m_k$ is typically fixed and independent of $n_k$;  see, for example, \citet{BenkoCFPC2007AOS} and \citet{jiao_filtrated_2024}. To better characterize the inherent infinite-dimensional nature of functional data, we instead allow $m_k$ to diverge as $n_k$ grows. Finally, let $\hs^{[k]}_0$ denote the subspace spanned by the leading $m_k$ eigenfunctions of source $k$. 

As shown in \eqref{eq:Mercer_thm}, the covariance structure is determined by its eigenfunctions, which capture variability patterns, and the corresponding eigenvalues, which quantify their magnitudes. Ideally, if the sample sizes and the number of observations per subject for different sources are roughly the same, the number of selected eigenfunctions is about the same. Furthermore, if $\phi_{\nu}^{[k]} = \phi_{\nu}$ for all $k \in \K$ and $\nu \in \N_{+}$, the shared subspace can be obtained from the leading eigenfunctions of the averaged sample covariance operators. In practice, however, these assumptions are often too restrictive. Real-world multi-source datasets may exhibit heterogeneity in covariance structures or imbalanced sample sizes, which introduces additional challenges for data integration in the following ways:
\begin{itemize}
    \item The number of eigenfunctions that can be consistently estimated depends on the sample size and the average of the number of observations per subject. Consequently, $m_k$ may vary across sources when $n_k$ or the average $N_i^{[k]}$ differ substantially.
    \item For some source $k$, the space spanned by its leading $m_k$ eigenfunctions can be decomposed into a shared subspace and a source-specific subspace. If the latter is not properly isolated, it may introduce bias into the recovery of the shared subspace.
    \item In some sources, the eigenfunctions associated with the shared subspace may correspond to relatively small eigenvalues, whereas source-specific eigenfunctions may be associated with larger eigenvalues. This imbalance can substantially degrade the performance of naive averaging approaches.
\end{itemize}

To integrate information across sources and address these challenges, we assume that there is a shared subspace $\hs_s$ with $\dim(\hs_s) = m_s$, spanned by the same eigenfunctions across all sources. Note that, we allow $m_s$ to diverge as the sample sizes increase. Thus, the shared structure can be formalized as follows: Let $\I_s^{[k]}$ contain the indices of the eigenfunctions that span $\hs_s$ for source $k \in \K$. Then by \eqref{eq:KL_expansion}, we rewrite $X^{[k]}(t)$ as
\begin{equation*}\label{eq:KL_expansion_with_shared_struc}
    X^{[k]}(t) = \sum_{\nu \in \I_s^{[k]}}\xi_\nu^{[k]}\phi_{\nu}^{[k]}(t) + \sum_{\nu \notin \I_s^{[k]}}\xi_\nu^{[k]}\phi_{\nu}^{[k]}(t) , ~~t \in \T.
\end{equation*}
Equivalently, for each $k\in\K$
$$\hs_{s}^{[k]} \triangleq \left\{f = \sum_{\nu = 1}^{m_s}\alpha_\nu \phi_{\pi_k(\nu)} \mid \alpha_1, \alpha_2, \ldots, \alpha_{m_s} \in \R \right\} = \hs_s,$$
where $\pi_k : \{1, 2, \ldots, m_s\} \to \I_s^{[k]}$ is a bijective function. This shared structure is similar to that defined in \cite{jiao_filtrated_2024}. Accordingly, define the projection operators
\begin{equation} \label{eq:share_private_space}
\bP_s^{[k]} = \sum_{\nu \in \I_s^{[k]}}\phi_\nu^{[k]}\otimes\phi_\nu^{[k]},\quad\bP_p^{[k]} = \sum_{\nu \notin \I_s^{[k]},~\nu \leq m_k}\phi_\nu^{[k]}\otimes\phi_\nu^{[k]},
\end{equation}
where $\bP_s^{[k]}$ projects onto the shared subspace and $\bP_p^{[k]}$ onto the source-specific subspace.

The following toy example illustrates the shared structure.
\begin{example}\label{example:Fourier_rotation}
Consider the Fourier basis functions:
\begin{singlespace}
    \begin{equation*}
    \phi_1(t) = 1,\quad
    \phi_{\nu}(t) =
    \begin{cases}
        \sqrt{2}\sin(\nu \pi t) \text{ if }\nu \geq 2 \text{ and } \nu \text{ is even},\\
        \sqrt{2}\cos\{(\nu-1) \pi t\} \text{ if }\nu \geq 2 \text{ and } \nu \text{ is odd.}
    \end{cases}
\end{equation*}
\end{singlespace}
Consider two sources with covariance operators $\bG^{[k]} = \sum_{\nu = 1}^{\infty}\lambda^{[k]}_\nu \phi_{\nu}^{[k]}\otimes\phi_{\nu}^{[k]}$ for $k = 1, 2.$ For source 1, set $\lambda_{\nu}^{[1]} = \nu^{-2}$ and $\phi_\nu^{[1]} = \phi_\nu$ for $\nu \in \N_+$. For source 2, we assume $\lambda_2^{[2]} = \lambda_{3}^{[1]}$, $\lambda_3^{[2]} = \lambda_{2}^{[1]}$, and $\lambda_\nu^{[2]} = \lambda_{\nu}^{[1]}$ for all other $\nu \in \N_+$. Define the eigenfunctions as
$$
    \phi_1^{[2]} = 2^{-1/2}(\phi_4 + \phi_5), \quad 
    \phi_2^{[2]} = \phi_2, \quad 
    \phi_3^{[2]} = \phi_3, \quad 
    \phi_4^{[2]} = \phi_1,
$$
and, for $\nu \geq 5$,
$$
    \phi_\nu^{[2]} = 2^{-1/2}(\phi_{2\nu-4} + \phi_{2\nu-3}).
$$
If $m_1 = m_2 = 3$, then the shared dimension is $m_s = 2$. In this case, $\pi_1(\nu) = \nu$ for $\nu \leq m_s$, while $\pi_2(1) = 3$ and $\pi_2(2) = 2$. 
\end{example}
\begin{remark}
In this construction, the eigenfunction associated with the largest eigenvalue always corresponds to the source-specific variation, whereas the shared subspace lies in the tail and is associated with relatively small eigenvalues. Consequently, weighting sample covariance operators and extracting leading eigenfunctions to identify the shared subspace may still perform poorly regardless of the choice of weights.
\end{remark}

\subsection{Multi-source FPCA}\label{sec:multiFPCA}
To identify the shared structure across multiple sources and generalize the standard FPCA, we consider focusing exclusively on the eigenfunctions while discarding the eigenvalues during the integration step. If source-specific variation is of further interest, analysis can be refined within the source of interest based on the estimated shared structure.

Motivated by the above idea and Example \ref{example:Fourier_rotation}, we propose the following projection-based estimator:
\begin{equation}\label{eq:Pw_estimator}
    \what{\bP}_w = n_t^{-1}\sum_{k \in \K}n_k\wtilde{\bP}^{[k]},
\end{equation}
where $n_t = \sum_{k\in\K}n_k$ and $\wtilde{\bP}^{[k]} = \sum_{\nu=1}^{m_k}\tilde{\phi}_\nu^{[k]}\otimes\tilde{\phi}_\nu^{[k]}$. Here $\tilde{\phi}_\nu^{[k]}$ is the $\nu$th eigenfunction estimated from the smoothed covariance function of the $k$th source. Note that, consistency of $\tphi_\nu^{[k]}$ implies the consistency of $\wtilde{\bP}^{[k]}$. 

The target of interest is the leading $m_s$ eigenfunctions of $\what{\bP}_w$, or equivalently, the subspace they span. Let $\what{\bP}_s$ denote the projection operator onto this estimated shared subspace. Then, it follows from \eqref{eq:PCA_projection_expression_alternative} that
\begin{align*}
    \what{\bP}_s 
    &= \underset{\bP \in \P_{m_s}}{\amax}~\frac{1}{n_t}\sum_{k \in \K}n_k\innerproductHS{\wtilde{\bP}^{[k]}}{\bP}\\[3pt]
    & = \underset{\bP \in \P_{m_s}}{\amin}~\frac{1}{n_t}\sum_{k \in \K}n_k\norm{\wtilde{\bP}^{[k]} - \bP}_{\text{HS}}^2,
\end{align*}
where $\P_{m_s}$ denotes the collection of rank-$m_s$ projection operators in $\hs$. Therefore, the projection operator of the shared structure minimizes the weighted average of Hilbert–Schmidt distances between $\wtilde{\bP}^{[k]}$ and $\bP$ across all sources.

\begin{figure}[!ht]
    \centering
    \includegraphics[width=0.95\textwidth, height = 20em]{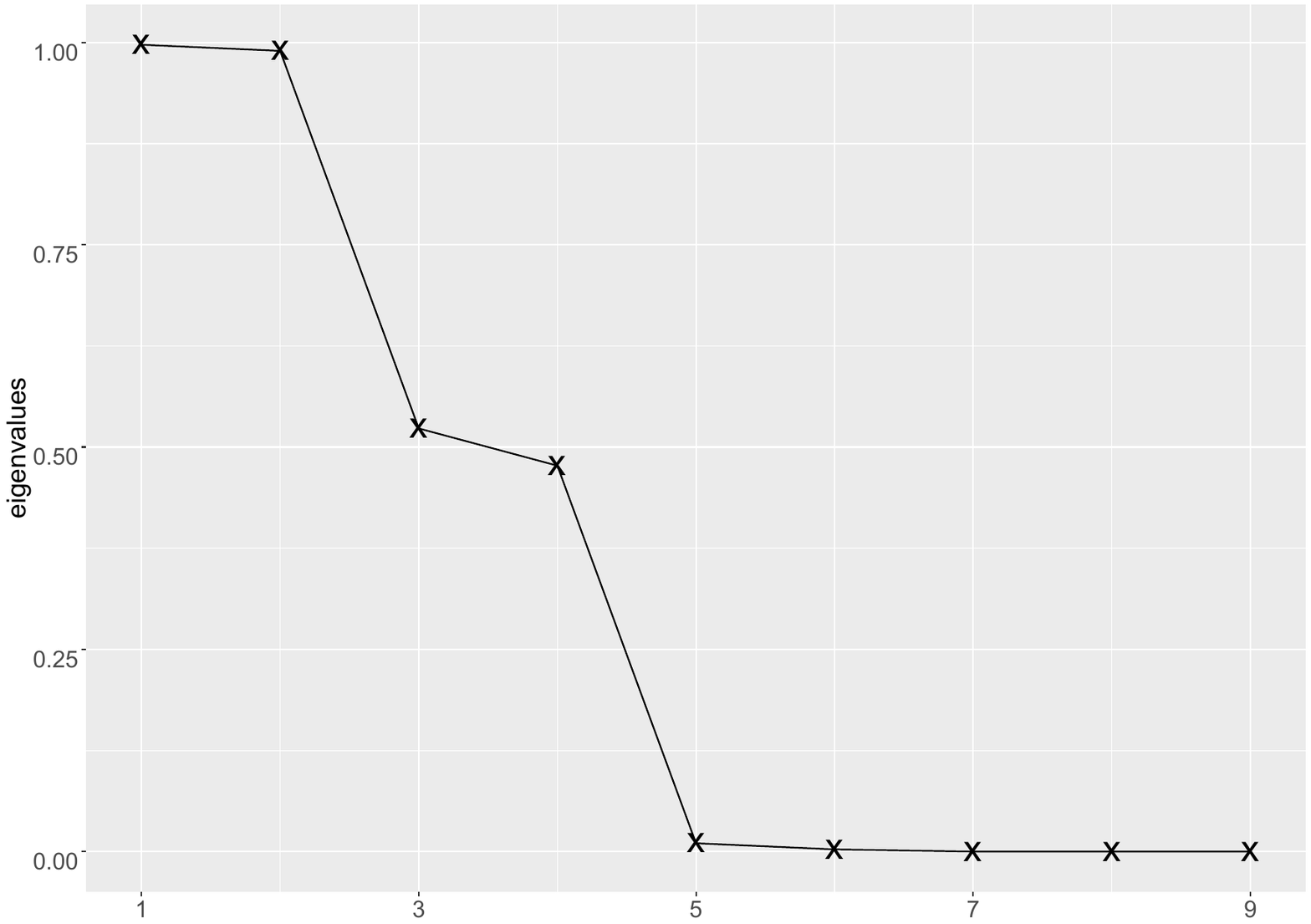}
    \caption{Scree plot of $\what{\bP}_w$ based on Example \ref{example:Fourier_rotation} with $m_1 = m_2 = 3$. Data are generated from two sources, each with 200 subjects and 25 observations per subject.}
    \label{fig:screeplot_example1}
\end{figure}

Note that at the population level, 
\begin{equation}\label{eq:Pw}
    \bP_w = n_t^{-1}\sum_{k \in \K}n_k\bP^{[k]} = \bP_s + \frac{1}{n_t}\sum_{k\in \K}n_k\bP_p^{[k]},
\end{equation}
which implies that the eigenvalues corresponding to the shared subspace are always equal to one. While the largest eigenvalue of $\sum_{k \in \K} n_k \bP_p^{[k]} / n_t$ is strictly smaller than one. Since $\wtilde{\bP}^{[k]}$ is consistent, $\what{\bP}_w$ is also consistent. Consequently, the rank of the shared subspace can be identified from the scree plot of $\what{\bP}_w$; see Figure \ref{fig:screeplot_example1} for illustration.
Further discussions of the eigengap between the shared subspace and $\sum_{k \in \K} n_k \bP_p^{[k]} / n_t$ are provided in Remark \ref{remark:eigengap} and Example \ref{example:eigengap}. Let $(\hat{\theta}_\nu, \hat{\psi}_\nu)$ denote the $\nu$th eigen-pair of $\what{\bP}_w$. Therefore, we have
\begin{equation}\label{eq:shared_projection_operator}
  \what{\bP}_s = \sum_{\nu=1}^{m_s}\hat{\psi}_\nu \otimes \hat{\psi}_\nu.
\end{equation}

If source-specific variation for source $k$ is of interest and $m_k>m_s$, the corresponding projector $\bP_p^{[k]}$, which is defined in \eqref{eq:share_private_space}, can be estimated using only the $k$th source data after obtaining $\what{\bP}_s$. The space associated with $\what{\bP}_s$ can be viewed as an average of the shared subspace across $K$ sources. Consequently, the basis functions generating $\what{\bP}_s$ are not necessarily orthogonal to those generating the source-specific variation. To preserve orthogonality, we need to define a projection operator $\what{\bP}_d$ that projects functions onto $\what{\hs}^{[k]}_d = \wtilde{\hs}_0^{[k]} \ominus \what{\hs}_s$, where $\wtilde{\hs}_0^{[k]}$ and $\what{\hs}_s$ are two subspaces associated with $\wtilde{\bP}^{[k]}$ and $\what{\bP}_s$, respectively. Define $\what{\bT}^{[k]} = \wtilde{\bP}^{[k]}(\idfun - \what{\bP}_s)\wtilde{\bP}^{[k]}$, where $\idfun$ is the identity operator that maps any functions in $\hs$ to itself. Denote by $\what{\bP}_d$ the projection operator generated by the eigenfunctions of $\what{\bT}^{[k]}$ associated with the eigenvalue one. Note that $\what{\bT}^{[k]}(f) = f$ for any $f \in \what{\hs}^{[k]}_d$, which implies $\what{\bP}_d (f) = f.$ The construction of $\what{\bP}_d$ implicitly forces any function not in $\wtilde{\hs}_0^{[k]} \cap \what{\hs}_s^{\perp}$ to be zero as we set these eigenvalues to be $0$. Thus, $\what{\bP}_d(f) = 0$ for any $f \in \hs \ominus \what{\hs}^{[k]}_d$. To ensure the eigenfunctions in the source-specific subspace are identifiable, we then project $\wtilde{\bG}^{[k]}$, the estimated covariance function for the $k$th source, onto $\what{\bP}_d^{[k]}$. The source-specific subspace $\what{\bP}_p^{[k]}$ is then represented by the leading $m_k - m_s$ eigenfunctions of $$\what{\bG}_p^{[k]} = \what{\bP}_d^{[k]}\wtilde{\bG}^{[k]}\what{\bP}_d^{[k]}.$$ The refined estimator of projection operator for the $k$th source is given by $\what{\bP}^{[k]} = \what{\bP}_p^{[k]} + \what{\bP}_s$. The entire procedure is summarized in Algorithm \ref{algo:multiFPCA}.

\begin{algorithm}[!ht]
  \caption{Multi-source FPCA}
  \vspace{1em}
  \begin{enumerate}
    \item For each source $k$, estimate the covariance function $\wtilde{G}^{[k]}(t,s)$ and eigenpairs $\{(\tilde{\lambda}_{\nu}^{[k]}, \tilde{\phi}_{\nu}^{[k]}) \mid \nu = 1, 2, \ldots, m_k\}$ using the method detailed in Section \ref{sec:standard_FPCA}. Moreover, compute $\wtilde{\bP}^{[k]} = \sum_{\nu = 1}^{m_k} \tilde{\phi}_{\nu}^{[k]} \otimes \tilde{\phi}_{\nu}^{[k]}$.
    \item Construct the weighted average projection operator $\what{\bP}_w$ as in equation \eqref{eq:Pw_estimator}, and perform eigendecomposition of its discretized version to obtain eigenpairs $(\hat{\theta}_\nu, \hat{\psi}_\nu)$.
    \item Identify the rank $m_s$ of $\hs_s$ based on the eigenvalues of $\what{\bP}_w$. The projection operator of the shared subspace is determined by using equation \eqref{eq:shared_projection_operator}.
    \item[4.1] (optional) If source-specific variation for source $k$ is of interest and $m_k > m_s$, obtain $\what{\bP}_d^{[k]}$ by acquiring the eigenspace of $\wtilde{\bP}^{[k]}(\idfun - \what{\bP}_s)\wtilde{\bP}^{[k]}$ associated with the eigenvalue equals one. Then, we re-project $\wtilde{\bG}^{[k]}$ onto the space associated with $\what{\bP}_d^{[k]}$ to obtain $\what{\bG}_p^{[k]}=\what{\bP}_d^{[k]}\wtilde{\bG}^{[k]}\what{\bP}_d^{[k]}$.
    \item[4.2] Extract the leading $m_k - m_s$ eigenfunctions of $\what{\bG}_p^{[k]}$ to estimate the source-specific subspace and its associated projection operator $\what{\bP}_p^{[k]}$.
\end{enumerate}
\label{algo:multiFPCA}
\end{algorithm}

\section{Statistical Theory}\label{sec:theory}
This section is devoted to the theoretical justification of the proposed method. The shared structure is formalized in Section \ref{sec:shared_structure}. In what follows, let $\normOP{\bA} = \sup_{f \in \hs} \norm{\bA(f)}_2 / \norm{f}_2$ denote the operator norm of $\bA$. For a compact operator, the operator norm equals its largest eigenvalue. For any continuous function $f$ defined on $[0, 1]$, $\|f\|_{\infty} = \sup_{t \in [0, 1]} |f(t)|$. 
Let $a_n \lesssim b_n$ denote $a_n \leq C b_n$ for all $n \in \mathbb{N}^+$ with some positive constant $C$, and $\gtrsim$ is defined similarly. Lastly, $a_n \asymp b_n$ if $a_n \lesssim b_n$ and $b_n \lesssim a_n$. We need the following assumptions to develop theoretical properties for the proposed estimator.
\begin{assp}\label{assp:shared_space}
The shared space $\hs_s$ is spanned by the same basis functions across all sources. Furthermore, we assume $m_s \leq \min_{k \in \K} m_k.$
\end{assp}
\begin{assp}\label{assp:private_space}
For the source-specific subspaces, we assume
\begin{equation*}
   \normOP[\bigg]{\frac{1}{n_t}\sum_{k \in \K}n_k \bP_p^{[k]}} \leq 1 - d,
\end{equation*}
where $d > 0$ is a constant that is independent of sample sizes.
\end{assp}
\begin{remark}\label{remark:eigengap}
Note that the largest eigenvalue of $\bP_w$ defined in \eqref{eq:Pw} satisfies $\lambda_1(P_w) = 1$. Assumption \ref{assp:private_space} essentially provides a lower bound on the discrepancy between the largest two eigenvalues of $P_w$. The magnitude of $1 - d$ characterizes the degree of similarity among the source-specific subspaces across sources. A smaller $d$ indicates greater challenges in estimating the shared space. In particular, if $d = 0$, then an eigenfunction is shared by all sources, and by Assumption \ref{assp:shared_space}, this eigenfunction must belong to $\hs_s$.
\end{remark}
\begin{example}\label{example:eigengap}
Let us revisit Example \ref{example:Fourier_rotation} and further assume $n_1 = n_2$. It is easy to see that $\norm[\Big]{2^{-1}\sum_{k \in \K}\bP_p^{[k]}}_{\text{op}} = 1/2$, and $d = 1/2$. If instead, we let $\phi^{[2]}_1= 2^{-1/2}(\phi_1 + \phi_4)$ and $\phi_4^{[2]} = \phi_5$ while keeping $m_1 = m_2 = 3$, then $\norm{2^{-1}\sum_{k \in K}\bP_p^{[k]}}_{\text{op}} = 1/2 + 2^{1/2}/4$, and $d = 1/2 - 2^{1/2}/4$. Thus, $d$ is smaller in second case as $\phi_1^{[2]}$ is partially aligned with $\phi_1^{[1]} = \phi_1$, indicating an overlap between two source-specific subspaces. In contrast, if $\phi_1^{[2]}$ is completely aligned with $\phi_1^{[1]}$, then $m_s = 3$ and this eigenfunction belongs to the shared space. 
\end{example}

Assumptions \ref{assp:scores_4th_moment} - \ref{assp:random_design} should be understood as they hold for any $k \in \K$. For brevity, we omit the superscript $[k]$ or $k$. These assumptions ensure that the covariance function and eigenpairs are well estimated for each source. 
\begin{assp} \label{assp:scores_4th_moment}
    There exists a positive constant $c_0$ such that $\E\xi_\nu^4 \leq c_0(\lambda_\nu)^2$.
\end{assp}

\begin{assp}\label{assp:partial_derivative_cov}
    The second order partial derivatives of $G(s,t)$, $\partial G(s,t)/\partial t\partial s$, $\partial G(s,t)/\partial s^2$, and $\partial G(s,t)/\partial t^2$, are all bounded on $\T \times \T$.
\end{assp}

\begin{assp}\label{assp:eigenvalue_decay_rate}
    The eigenvalues of $\bG$ follow a polynomial decay rate. That is, there exists a positive constant $a > 1$ such that, for any $\nu \in \N_+$,
    \begin{equation*}
        \nu^{-a} \gtrsim \lambda_{\nu} \gtrsim \lambda_{\nu + 1} \gtrsim \nu^{-(a+1)}
    \end{equation*}
\end{assp}

\begin{assp} \label{assp:smooth_cond_eigenfunc}
    Suppose $\norm{\phi_\nu}_{\infty} = O(1)$ for $\nu \in \N_+$, and there exists a positive constant $c$
    \begin{equation*}
        \norm{\phi_\nu^{(\ell)}}_\infty \lesssim \nu^{c/2}\norm{\phi_\nu^{(\ell - 1)}}_\infty\quad \text{for } \ell = 1, 2,
    \end{equation*}
    where $f^{(\ell)}$ denotes the $\ell$th derivative of $f$.
\end{assp}

\begin{assp}\label{assp:exp_tails_X_and_eps}
    $\E(\norm{X}^{2\alpha}_\infty)$ and $\E(|\eps|^{2\alpha})$ are bounded for some $\alpha > 3$.
\end{assp}

\begin{assp}\label{assp:random_design}
    The observation times $T_{ij}$ are independent of $X$ and $\eps$. Moreover, $T_{ij}$'s are iid with a density function bounded away from zero and infinity on $\T$.
\end{assp}

Assumptions \ref{assp:scores_4th_moment} and \ref{assp:partial_derivative_cov} are widely adopted in the functional data literature when smoothing methods are employed; see \cite{yao_functional_2005} and \cite{cai_optimal_2011} for example. Assumption \ref{assp:eigenvalue_decay_rate} considers polynomial decay rates for eigenvalues, where the resulting lower bounds of eigengaps facilitate a theoretical framework for the estimated functional principal components \citep{hall_methodology_2007, zhou_functional_2023}. Assumption \ref{assp:smooth_cond_eigenfunc} was considered in \cite{zhou_theory_2024} to control the smoothness of eigenfunctions and their derivatives. In particular, it is easy to see that $\norm{\phi^{(2)}_\nu}_\infty \lesssim \nu^c$, which implies that higher order derivatives can grow up to a polynomial rate with respect to the index $\nu$ of the eigenfunction. The moments condition in Assumption \ref{assp:exp_tails_X_and_eps} ensures the uniform convergences of the estimated covariance function, which can be found, for example, in \cite{li_uniform_2010} and \cite{zhang_sparse_2016}.

We first present the convergence rate of $\what{\bP}_s$ in terms of the operator norm. Note that the projection operator is the aggregation of the projection operator for each source. Thus, Theorems \ref{thm:shared_space_thm} and \ref{thm:private_space_thm} are based on the FPCA theory for a single source.

\begin{thm}\label{thm:shared_space_thm}
Suppose Assumptions \ref{assp:shared_space} and \ref{assp:private_space} hold. For each source $k$, suppose Assumptions \ref{assp:scores_4th_moment} - \ref{assp:random_design} hold with respective constants, and denote $$\bar{N}_2^{[k]} = \left\{n_k^{-1}\sum_{i=1}^{n_k}(N_i^{[k]})^{-2}\right\}^{-1/2}.$$
 For any $k \in \K$, we further assume $m_k^{2a_k+2}/n_k = o(1)$,  $\bar{N}_2^{[k]} \gtrsim m_k$, and for $\nu \leq m_k$, 
$$\frac{m_k^{2a_k + 2}}{n_k(\bar{N}_2^{[k]})^{2}h_k^2(\nu)} = o(1), ~h^{4}_k(\nu)\max\{m_k^{2a_k+2c_k}, m_k^{4a_k}\log(n_k)\} = o(1),$$ 
where $$h_k(\nu) = (n_k\bar{N}_2^{[k]})^{-1/5}\nu^{(a_k - 2c_k-2)/5}(1 + \nu^{a_k}/\bar{N}_2^{[k]})^{1/5}.$$ 
Under these assumptions, we have 
$$\normOP{\what{\bP}_s - \bP_s} = O_p(\kappa_{n_t}).$$
Here, 
\begin{itemize}
    \item When $\bar{N}_2^{[k]} \gtrsim m_k^{a_k}$, we further assume $c_k + 2 < 4a_k$  for all $k \in \K$. Then,
    \begin{equation*}
        \kappa_{n_t} = \frac{\left(\sum_{k \in \K}m_k^4\right)^{1/2}}{n_t^{1/2}} + \frac{\pbrac[\Big]{\sum_{k \in \K}m_k^{c_k/2+3}}^{2/5}}{n_t^{2/5}}.
    \end{equation*}
    If $\bar{N}_2^{[k]} \gtrsim n_k^{1/4}m_k^{a_k + c_k/2 -2}$ for all $k \in \K$, then
    \begin{equation*}
        \kappa_{n_t} = \frac{\left(\sum_{k \in \K}m_k^4\right)^{1/2}}{n_t^{1/2}}.
    \end{equation*}
    \item When $\bar{N}_2^{[k]} \asymp m_k$ for all $k \in \K$, we further assume $\max_{k \in \K} 2(a_{k}+1)/(a_k + c_{k}/2 + 1) > 1$. Then,
    \begin{equation*}
        \kappa_{n_t} = \frac{\pbrac[\Big]{\sum_{k \in \K}m_k^{2a_k+2}}^{1/2}}{n_t^{1/2}} + \frac{\pbrac[\Big]{\sum_{k \in \K}m_k^{2a_k+c_k/2 + 1}}^{2/5}}{n_t^{2/5}}.
\end{equation*}
\end{itemize}
\end{thm}
\begin{remark}
The additional conditions on $a_k$ and $c_k$ arise from aggregating estimated eigenfunctions within the shared space, which are very mild. For example, standard bases used in FDA, including Fourier, Legendre, and wavelet, satisfy Assumption \ref{assp:smooth_cond_eigenfunc} with $c_k = 2$. The polynomial decay rate for eigenvalues in Assumption \ref{assp:eigenvalue_decay_rate} normally assumes $a_k > 1$. Consequently, both $c_k + 2 < a_k$ and $\max_{k \in \K} 2(a_{k}+1)/(a_k + c_{k}/2 + 1) > 1$ hold.
\end{remark}
\begin{remark}\label{remark:shared_space_rate_comparison}
To illustrate the efficiency gain from data integration, consider the following example. For simplicity, we focus on the dense case, where $\kappa_{n_t} = \left(\sum_{k \in \K} m_k^4 \right)^{1/2} / n_t^{1/2}$. Suppose there exists a source $k_0$ such that $n_{k_0} \asymp n_t$ and $n_k / n_{k_0} = o(1)$ for all $k \neq k_0$. When $m_k \asymp m$ for all $k$, we have $\kappa_{n_t} \asymp m^2 / n_{k_0}^{1/2}$. In contrast, if only source $k$ is available for some $k \neq k_0$, the convergence rate for estimating the projection operator of the shared subspace is $O_p[(m_sm) / n_k^{1/2}]$; see Proposition 1 in the supplementary material. This rate is based on the oracle setting where the true indices of the shared eigenfunctions are known for source $k$. The ratio between the two convergence rates is $(mn_k)^{1/2} / (m_sn_{k_0})$. For example, if $m_s \asymp m$, the ratio is $o(1)$. This demonstrates the improvement in estimation efficiency achieved through data integration. This theoretical result aligns with the numerical findings reported in Table \ref{tb:shared_space_comparison} in Section \ref{sec:simulation}.
\end{remark}
\begin{thm}\label{thm:private_space_thm}
Under Assumptions for Theorem \ref{thm:shared_space_thm}, if $m_k > m_s$, then the estimator of the $k$th source-specific subspace satisfies 
$$\normOP[\Big]{\what{\bP}_p^{[k]} - \bP_p^{[k]}} = O_p(\zeta_{n_k}).$$
Here,
\begin{itemize}
    \item When $\bar{N}_2^{[k]} \gtrsim m_k^{a_k}$, 
    \begin{equation*}
        \zeta_{n_k} = \frac{(m_k - m_s)^{2}}{n_k^{1/2}} + \left\{\frac{(m_k - m_s)^{c_k/2 + 3}}{n_k}\pbrac[\Big]{\frac{m_k - m_s}{m_k}}^{a_k}\right\}^{2/5}.
    \end{equation*}
    If we further assume $\bar{N}_2^{[k]} \gtrsim n_k^{1/4}m_k^{a_k + c_k/2 -2}$, then
    \begin{equation*}
        \zeta_{n_t} = \frac{(m_k - m_s)^{2}}{n_k^{1/2}}.
    \end{equation*}
    \item When $\bar{N}_2^{[k]} \asymp m_k$,
    \begin{equation*}
        \zeta_{n_k} = \frac{(m_k - m_s)^{a_k + 1}}{n_k^{1/2}} + \frac{(m_k - m_s)^{(4a_k + c_k + 2)/5}}{n_k^{2/5}}.
\end{equation*}
\end{itemize}
\end{thm}

\begin{remark}\label{remark:private_space_rate_comparison}
Let the source-specific subspace for source $k_0$ be denoted by $\hs_d^{[k_0]} = \hs_0^{[k_0]} \ominus \hs_s$. When $m_{k_0} > m_s$, it is not surprising that the convergence rate in Theorem \ref{thm:private_space_thm} is of the same order as that of a single-source estimator, unless $\dim(\hs_d^{[k_0]}) \ll \dim(\hs_s)$. 
In particular, we focus on the dense case discussed in Remark \ref{remark:shared_space_rate_comparison}, and suppose $m_k \asymp m$ for all $k \in \K$. In the oracle setting where the true indices of the eigenfunctions spanning $\hs_d^{[k_0]}$ are known, the convergence rate for estimating the projection operator of the source-specific subspace is $O_p[(m-m_s)m/n_{k_0}^{1/2}]$ based on Proposition 1 in the supplementary material,
while Theorem \ref{thm:private_space_thm} gives $\zeta_{n_{k_0}} = (m - m_s)^2/n_{k_0}^{1/2}$. Thus, the ratio between the two convergence rates is $1/(1 + r)$, where $r = \dim(\hs_s)/\dim(\hs_d^{[k_0]})$. This result implies that if $\dim(\hs_d^{[k_0]}) \asymp \dim(\hs_s)$, the efficiency gain is asymptotically constant; however, when $\dim(\hs_d^{[k_0]}) \ll \dim(\hs_s)$, the gain becomes substantial. 

For example, in the extreme case where $\dim(\hs_d^{[k_0]}) = 1$ and $m_s \asymp m \to \infty$ as $n_k \to \infty$ for all $k \in \K$, the estimated subspace $\widehat{\hs}_d^{[k_0]}$ must be orthogonal to the estimated shared subspace, and accurate estimation of $\hs_s$ markedly improves the recovery of $\hs_d^{[k_0]}$. Conversely, if $\dim(\hs_s) = 1$ but $\dim(\hs_d^{[k_0]}) \to \infty$ as $n_k \to \infty$, improvements in estimating $\hs_s$ have negligible effect on the estimation of $\hs_d^{[k_0]}$.
In summary, although the information for estimating the source-specific subspace resides solely within source $k_0$, the orthogonality constraint between $\hs_s$ and $\hs_0^{[k_0]} \ominus \hs_s$ can help enhance estimation efficiency.
\end{remark}

\section{Simulation}\label{sec:simulation}
In this section, we perform simulation studies to investigate the finite sample performance of the proposed method, denoted as Multi-FPCA. We consider three sources in total. To showcase the superior performance of the proposed method, we consider three different sample size settings: 1.) $n_1 = 50$ and $n_2 = n_3 = 200$; 2.) $n_1 = 100, n_2 = n_3 = 200$; 3.) $n_1 = 100, n_2 = n_3 = 400$. In addition, we consider the number of observations per subject $N_i = N \in \{10, 25, 50\}$, where $N=10$ and $50$ are referred to as sparse and dense functional data, respectively. The intermediate case, $N = 25$, represents a transition state between sparse and dense, hereby referred to as neither.

We generate data based on model \eqref{eq:model} and the Karhunen–Lo\`{e}ve expansion in \eqref{eq:KL_expansion}. Observation times $T_{ij}^{[k]}$ are uniformly generated in $[0, 1]$. For brevity, we assume the true mean function is zero. In terms of the covariance structure, we first denote the ordinary Fourier basis as follows:
\begin{singlespace}
    \begin{equation*}
    \phi_1(t) = 1,\quad
    \phi_{\nu}(t) =
    \begin{cases}
        \sqrt{2}\sin(\nu \pi t) \text{ if }\nu \geq 2 \text{ and } \nu \text{ is even},\\
        \sqrt{2}\cos\{(\nu-1) \pi t\} \text{ if }\nu \geq 2 \text{ and } \nu \text{ is odd.}
    \end{cases}
\end{equation*}
\end{singlespace}
The eigen-systems for the three sources are specified as follows:
\begin{enumerate}[label=Source \arabic*:, leftmargin=4.8em]
    \item Set $\lambda_1^{[1]} = 24, \lambda_2^{[1]} = 12, \lambda_3^{[1]} = 6$, and $\lambda_\nu = 0$ for $\nu \geq 4$. The eigenfunctions are $\phi_\nu^{[1]} = \phi_\nu$ for $\nu \in \N_{+}$. The variance of the measurement error, $\sigma_{\eps}^2$, is $0.49$.
    \item Set $\lambda_1^{[2]} = 12, \lambda_2^{[2]} = 10, \lambda_3^{[2]} = 8$, $\lambda_4^{[2]} = 4$, and $\lambda_\nu = 0$ for $\nu \geq 5$. As for the eigenfunctions, $\phi_1^{[2]} = \phi_3$, $\phi_2^{[2]} = (\sin\theta)\phi_4 + (\cos\theta)\phi_5$, $\phi_3^{[2]} = \phi_2$, $\phi_4^{[2]} = (\sin\theta)\phi_6 + (\cos\theta)\phi_7$ with $\theta = \pi/4$. The variance of the measurement error, $\sigma_{\eps}^2$, is $0.16$.
    \item Set $\lambda_1^{[3]} = 20, \lambda_2^{[3]} = 10, \lambda_3^{[3]} = 5$, and $\lambda_\nu = 0$ for $\nu \geq 4$. The eigenfunctions are $\phi_1^{[3]} = (\sin\omega)\phi_4 + (\cos\omega)\phi_5$, $\phi_2^{[3]} = \phi_2$, $\phi_3^{[3]} = \phi_3$ with $\omega = \pi/3$. The variance of the error, $\sigma_{\eps}^2$, is $0.16$.
\end{enumerate}
The above settings imply $m_s = 2$. 

To illustrate the benefit of integrating information, we compare Multi-FPCA with an estimator using only source 1 data. For the latter, we first estimate $\wtilde{\bP}^{[1]}$ with the method in Section \ref{sec:standard_FPCA}. We then compute $\wtilde{\bP}_s^{[1]}$ and $\wtilde{\bP}_p^{[1]}$ by providing the true indices of the shared eigenfunctions in the first source. In particular, $\wtilde{\bP}^{[1]}_s = \tilde{\phi}_2^{[1]}\otimes \tilde{\phi}_2^{[1]} + \tilde{\phi}_3^{[1]}\otimes \tilde{\phi}_3^{[1]}$. It is worth noting that the true indices are provided in computing $\wtilde{\bP}^{[1]}_s$ and $\wtilde{\bP}^{[1]}_p$, while the indices of the shared eigenfunctions are always estimated from the data for Multi-FPCA. The estimation procedure for the Multi-FPCA is detailed in Section \ref{sec:multiFPCA} and summarized in Algorithm \ref{algo:multiFPCA}. 500 independent simulation trials were run for each setting mentioned above.

To assess the performance of these methods, we consider the following metric:
\begin{equation*}
    \frac{1}{M}\sum_{i=1}^{M}\norm{\what{\bP}_i - \bP},
\end{equation*}
where $M = 500$ and $\norm{\cdot}$ denotes either the operator norm or Hilbert–Schmidt norm. Here $\what{\bP}_i$ denotes the generic estimate of the target (sub)space of interest, i.e., the entire space, the shared subspace or the source-specific subspace,
based on Multi-FPCA or using the source 1 data only from the $i$th simulation run, and $\bP$ is the corresponding true (sub)space. 

We report both the averaged norm differences and the associated standard error. The results of estimating the shared subspace and the source-specific subspace for source 1 are displayed in Table \ref{tb:shared_space_comparison} and Table \ref{tb:source_specific_space_comparison}, respectively. Table \ref{tb:entire_space_comparison} summarizes the results for estimating the entire space for source 1.

\begin{table}[!ht]
\centering
\caption{Mean and standard error (in parenthesis) comparison of the norm differences for the \textit{shared} subspace across 500 Monte Carlo runs between Multi-FPCA and using the source 1 data only.}
\vspace{0.5em}
\begin{tabular}{c c @{\hspace{1em}} c c c c}
  \hline
  \hline\\[-2.2ex]
  \multirow{2}{*}{$(n_1, n_2, n_3)$} & \multirow{2}{*}{$N$} & \multicolumn{2}{c}{Multi-FPCA} & \multicolumn{2}{c}{Source 1 only}\\[0.5ex]
  \cline{3-6}\\[-2.2ex]
  & & $\normOP{\cdot}$ & $\normHS{\cdot}$ & $\normOP{\cdot}$ & $\normHS{\cdot}$\\[0.5ex]
  \hline\\[-2.2ex]
  \multirow{3}{*}{$(50, 200, 200)$} & 10 &  0.188(0.06) & 0.294(0.09) & 0.306(0.10) & 0.475(0.15)\\
  & 25 & 0.117(0.03) & 0.188(0.04) & 0.251(0.09) & 0.388(0.13)\\
  & 50 &  0.089(0.02) & 0.144(0.03) & 0.247(0.11) & 0.377(0.14)\\[0.5ex]
  \hline\\[-2.2ex]
  \multirow{3}{*}{$(100, 200, 200)$} & 10 &  0.160(0.04) & 0.254(0.06) & 0.238(0.07) & 0.368(0.11)\\
  & 25 & 0.101(0.02) & 0.166(0.04) & 0.196(0.07) & 0.306(0.10)\\
  & 50 &  0.077(0.01) & 0.126(0.02) & 0.181(0.07) & 0.279(0.09)\\[0.5ex]
  \hline\\[-2.2ex]
  \multirow{3}{*}{$(100, 400, 400)$} & 10 & 0.142(0.05) & 0.224(0.06) & 0.237(0.07) & 0.368(0.10)\\
  & 25 &  0.088(0.02) & 0.142(0.03) & 0.189(0.06) & 0.296(0.09)\\
  & 50 &  0.067(0.02) & 0.108(0.02) & 0.176(0.06) & 0.272(0.08)\\[0.5ex]
  \hline
  \hline
\end{tabular}
\label{tb:shared_space_comparison}
\end{table}

\begin{table}[!ht]
\centering
\caption{Mean and standard error (in parenthesis) comparison of the norm differences for the \textit{source-specific} subspace across 500 Monte Carlo runs between Multi-FPCA and using the source 1 data only.}
\vspace{0.5em}
\begin{tabular}{c c @{\hspace{1em}} c c c c}
  \hline
  \hline\\[-2.2ex]
  \multirow{2}{*}{$(n_1, n_2, n_3)$} & \multirow{2}{*}{$N$} & \multicolumn{2}{c}{Multi-FPCA} & \multicolumn{2}{c}{Source 1 only}\\[0.5ex]
  \cline{3-6}\\[-2.2ex]
  & & $\normOP{\cdot}$ & $\normHS{\cdot}$ & $\normOP{\cdot}$ & $\normHS{\cdot}$\\[0.5ex]
  \hline\\[-2.2ex]
  \multirow{3}{*}{$(50, 200, 200)$} & 10 &  0.122(0.07) & 0.167(0.08) & 0.208(0.11) & 0.287(0.15)\\
  & 25 & 0.091(0.03) & 0.126(0.04) & 0.201(0.10) & 0.279(0.15)\\
  & 50 &  0.085(0.05) & 0.117(0.05) & 0.216(0.12) & 0.299(0.16)\\[0.5ex]
  \hline\\[-2.2ex]
  \multirow{3}{*}{$(100, 200, 200)$} & 10 &  0.103(0.07) & 0.142(0.08) & 0.152(0.09) & 0.209(0.11)\\
  & 25 & 0.078(0.02) & 0.109(0.03) & 0.153(0.07) & 0.212(0.1)\\
  & 50 &  0.067(0.02) & 0.093(0.02) & 0.153(0.08) & 0.212(0.11)\\[0.5ex]
  \hline\\[-2.2ex]
  \multirow{3}{*}{$(100, 400, 400)$} & 10 & 0.094(0.08) & 0.128(0.08) & 0.158(0.10) & 0.216(0.11)\\
  & 25 &  0.073(0.02) & 0.102(0.03) & 0.150(0.07) & 0.207(0.10)\\
  & 50 &  0.065(0.02) & 0.090(0.02) & 0.148(0.07) & 0.205(0.09)\\[0.5ex]
  \hline
  \hline
\end{tabular}
\label{tb:source_specific_space_comparison}
\end{table}

\begin{table}[!ht]
\centering
\caption{Mean and standard error (in parenthesis) comparison of the norm differences for the \textit{entire} space across 500 Monte Carlo runs between Multi-FPCA and using the source 1 data only.}
\vspace{0.5em}
\begin{tabular}{c c @{\hspace{1em}} c c c c}
  \hline
  \hline\\[-2.2ex]
  \multirow{2}{*}{$(n_1, n_2, n_3)$} & \multirow{2}{*}{$N$} & \multicolumn{2}{c}{Multi-FPCA} & \multicolumn{2}{c}{Source 1 only}\\[0.5ex]
  \cline{3-6}\\[-2.2ex]
  & & $\normOP{\cdot}$ & $\normHS{\cdot}$ & $\normOP{\cdot}$ & $\normHS{\cdot}$\\[0.5ex]
  \hline\\[-2.2ex]
  \multirow{3}{*}{$(50, 200, 200)$} & 10 &  0.191(0.08) & 0.307(0.10) & 0.261(0.11) & 0.403(0.14)\\
  & 25 & 0.126(0.03) & 0.213(0.05) & 0.182(0.05) & 0.296(0.08)\\
  & 50 &  0.103(0.05) & 0.176(0.05) & 0.152(0.06) & 0.257(0.07)\\[0.5ex]
  \hline\\[-2.2ex]
  \multirow{3}{*}{$(100, 200, 200)$} & 10 &  0.159(0.07) & 0.256(0.08) & 0.207(0.09) & 0.320(0.11)\\
  & 25 & 0.105(0.02) & 0.179(0.04) & 0.145(0.04) & 0.237(0.06)\\
  & 50 &  0.084(0.01) & 0.144(0.02) & 0.115(0.03) & 0.196(0.04)\\[0.5ex]
  \hline\\[-2.2ex]
  \multirow{3}{*}{$(100, 400, 400)$} & 10 & 0.149(0.08) & 0.239(0.09) & 0.207(0.09) & 0.320(0.11)\\
  & 25 &  0.097(0.02) & 0.162(0.03) & 0.139(0.04) & 0.229(0.06)\\
  & 50 &  0.078(0.02) & 0.132(0.02) & 0.116(0.03) & 0.196(0.04)\\[0.5ex]
  \hline
  \hline
\end{tabular}
\label{tb:entire_space_comparison}
\end{table}
Generally, the estimation accuracy improves for both methods as sample sizes or the number of observations per subject increases in terms of the norm differences. Tables \ref{tb:shared_space_comparison} - \ref{tb:entire_space_comparison} evidently illustrate the superior performance of Multi-FPCA. From Table \ref{tb:shared_space_comparison}, we find Multi-FPCA always outperforms the method using only single source data even if the true indices of the shared eigenfunctions are provided for the latter case. This confirms that Multi-FPCA effectively recovers the shared structure by leveraging information from additional sources, which also helps to reduce the estimation uncertainty. The improved shared-space estimation further enhances the recovery of the source-specific subspace as shown in Table \ref{tb:source_specific_space_comparison}. These two improved estimators,
$\what{\bP}_s$ and $\what{\bP}_p^{[1]}$, 
together lead to a better estimate for the entire subspace, as shown in Table \ref{tb:entire_space_comparison}. 

\section{Real Data Illustration}\label{sec:real_data}
Air pollution has become one of the major environmental concerns in China due to its rapid industrial expansion and urbanization over the past four decades \citep{guan2016impact}. Fine particulate matter (PM) with an aerodynamic diameter less than $\SI{2.5}{\micro\meter}$, often referred to as PM$_{2.5}$, is one of the main pollutants in multiple cities in China \citep{huang_high_2014}. Cardiovascular and respiratory diseases, and even lung cancer, are associated with long-term exposure to PM$_{2.5}$ \citep{PopeJAMA2002, hoek_long-term_2013, Lelieveld2015Nature}.

In this study, our objective is to identify the shared variation pattern of the daily PM$_{2.5}$ levels of five Chinese major cities: Beijing, Chengdu, Guangzhou, Shanghai, and Shenyang. These cities are located in economically developed regions, collectively contributing over 10\% of China’s Gross Domestic Product and approximately 6\% of its total population. The data were collected by U.S. diplomatic posts located in the five cities. The U.S. Embassy in Beijing started releasing hourly PM$_{2.5}$ readings in April 2008, followed by the consulates in Guangzhou, Shanghai, Chengdu, and Shenyang in November and December 2011, June 2012, and April 2013, respectively.

To begin, we stratify the data by human activity patterns and seasonal effects. Holidays and weekends are excluded to ensure that the selected days reflect homogeneous human activity patterns. Given the substantial influence of meteorological conditions on daily PM$_{2.5}$ levels, we analyze the data separately for winter months (November to February of the following year) and summer months (May to August). This partitioning accounts for the winter heating regulated by local governments in northern cities such as Beijing and Shenyang, where heating typically begins in November, following the approach of \citet{LiangJRSSA2015}. Finally, daily measurements are aggregated into weekly averages to mitigate temporal dependence.

\begin{figure}[!ht]
    \centering
    \includegraphics[width=\textwidth, height = 18em]{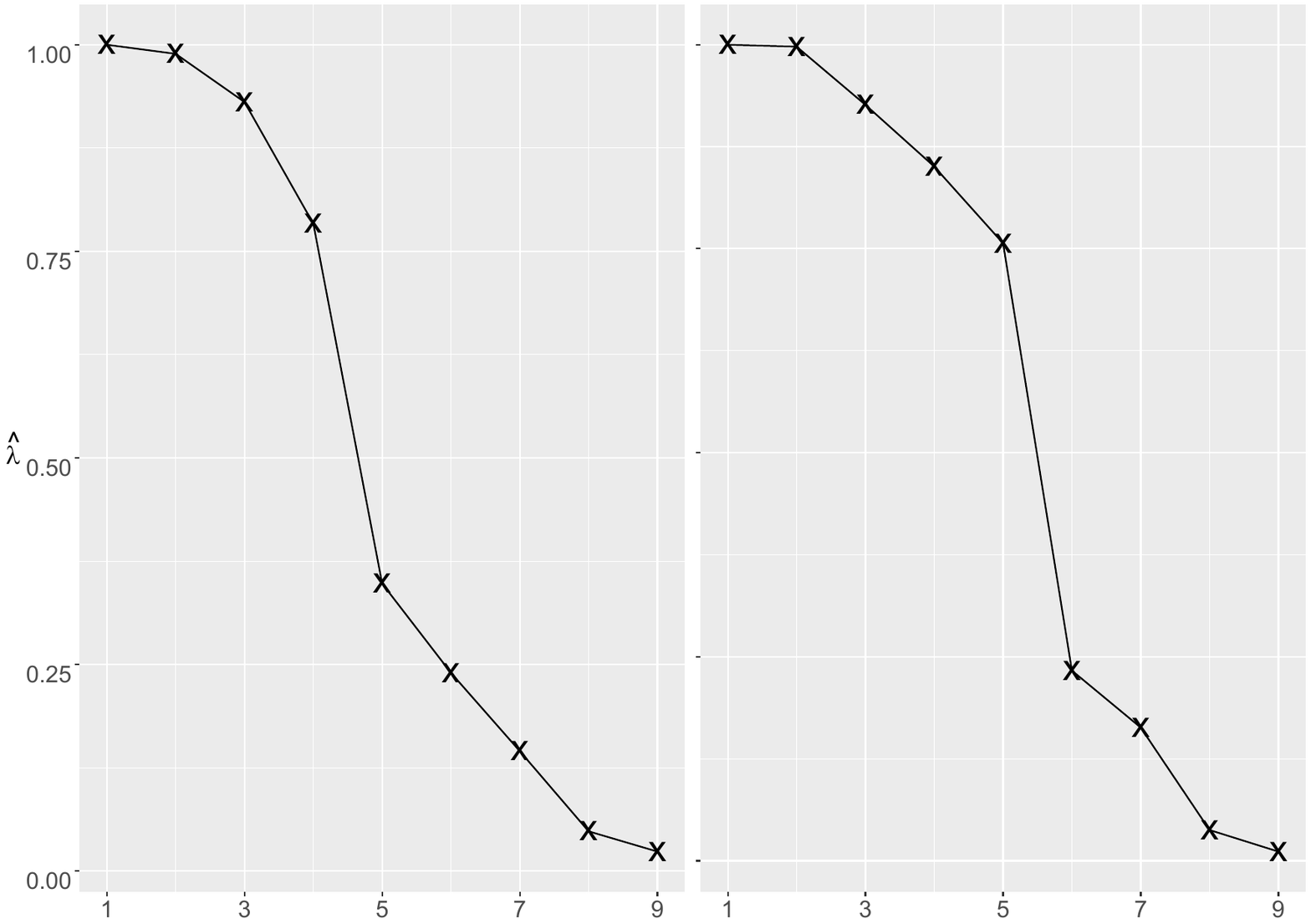}
    \caption{The left panel shows the estimated eigenvalues of $\what{\bP}_\text{winter}$ for five cities in the winter months, and the right panel demonstrates the estimated eigenvalues of $\what{\bP}_\text{summer}$ in the summer months. Both projected operators are calculated based on equation \eqref{eq:Pw_estimator}.}
    \label{fig:5_cities_eigenvals}
\end{figure}

The numbers of selected FPCs are 4, 4, 6, 4, 5 for Beijing, Chengdu, Guangzhou, Shanghai, and Shenyang, respectively, in winter, and 5, 6, 4, 5, 5 in summer. Figure \ref{fig:5_cities_eigenvals} displays the eigenvalues obtained from the weighted average of the projection operator for the winter and summer months. In general, the alignment of variability patterns across five cities is stronger in summer than in winter, suggesting human activities patterns are much more similar in summer. 

\begin{figure}[!ht]
    \centering
    \includegraphics[width=\textwidth, height = 18em]{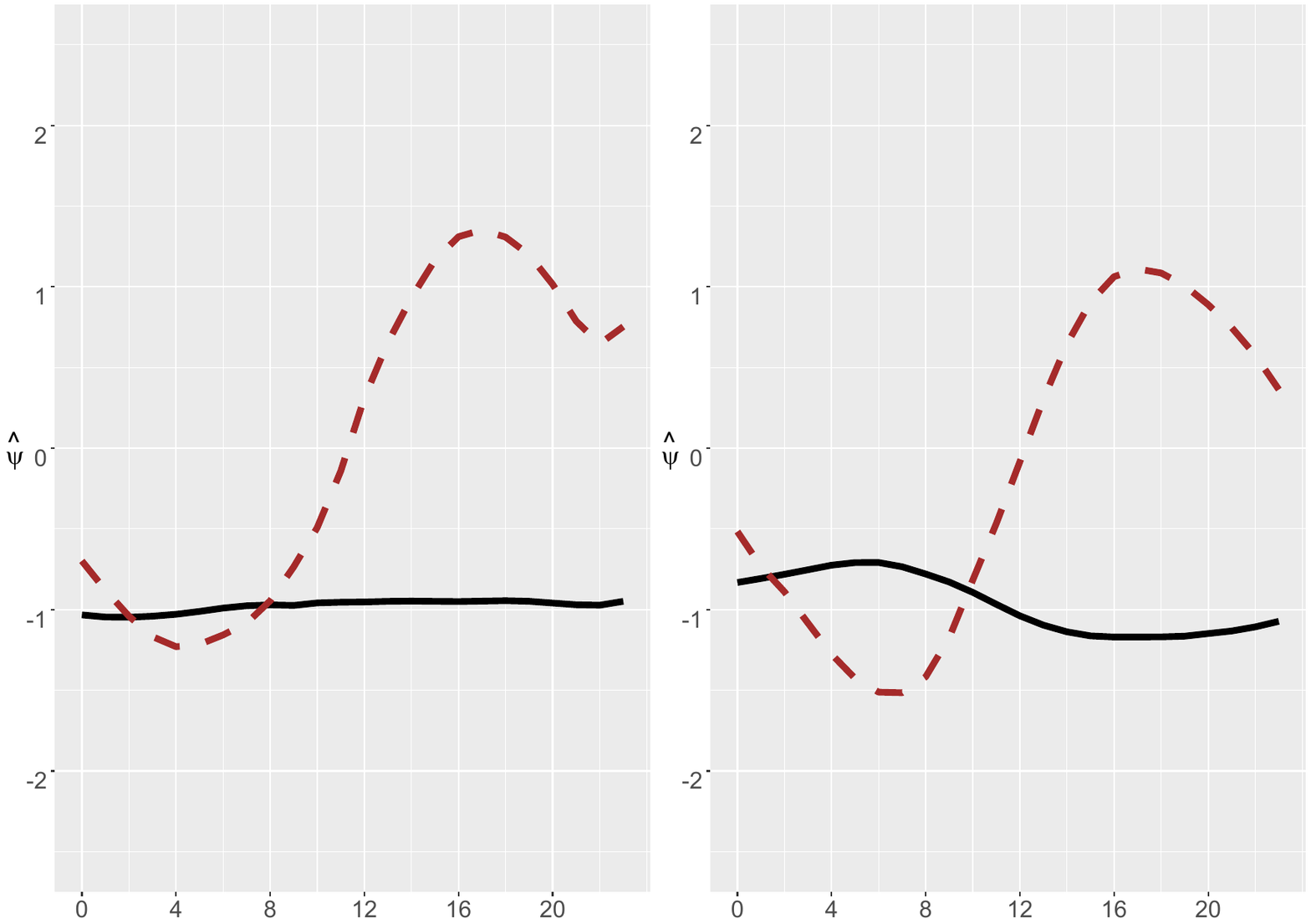}
    \caption{The left panel displays the leading two estimated eigenfunctions of $\what{\bP}_{\text{winter}}$, and the right panel presents those of $\what{\bP}_{\text{summer}}$. In both panels, the solid and dashed lines represent $\hpsi_1$ and $\hpsi_2$, respectively, obtained from the corresponding $\what{\bP}_s$.}
    \label{fig:5_cities_eigenfncs}
\end{figure}

The top two eigenvalues are close to one for both seasons, implying two dominant shared components. The associated eigenfunctions are shown in Figure \ref{fig:5_cities_eigenfncs}.
In both seasons, $\hpsi_2$ increases after the morning rush hour, peaks at approximately 4:30 PM, and declines slightly thereafter, suggesting that it captures afternoon traffic emissions. The shape of $\hpsi_1$ differs between winter and summer, likely reflecting meteorological effects. Overall, these findings indicate the existence of a shared structure in the variation pattern of daily PM$_{2.5}$ levels across the five cities, despite their substantial geographical and climatic diversity.

\section{Concluding Remarks}\label{sec:conclusions}
In this paper, we propose a novel procedure for performing FPCA across multiple data sources using projection operators. It adapts to both sparse and dense functional data. The primary goal is to recover the shared variation pattern across sources. Specifically, we estimate a projection operator from each source via local linear regression. Then we construct a weighted averaged projection operator  by aggregating the projection operator of each source. With the help of this estimator, we can unveil the variation pattern across sources even if the eigenvalues of the eigenfunctions spanned this subspace are relatively small within individual sources. When the source-specific variation is of interest, we further propose a re-projection method to estimate the source-specific subspace. We establish  asymptotic properties for both the shared space estimator and the source-specific variation estimator. Extensive numerical studies demonstrate the superior performance of the proposed methods. Lastly, we illustrate our method through analyzing the air pollutant data in five major cities in China.

\bibliographystyle{apalike}      
\bibliography{bib}   
\end{document}